\documentstyle[preprint,aps]{revtex}

\begin{document}
\title{Evidence for a transition from a bulk Meissner-state to a spontaneous vortex
phase in RuSr$_{2}$GdCu$_{2}$O$_{8}$ from DC magnetisation measurements}
\author{C. Bernhard$^{1)}$, J.L. Tallon$^{2)}$, E. Br\"{u}cher$^{1)}$, and R.K.
Kremer$^{1)}$}
\address{1) Max-Planck-Institut f\"{u}r Festk\"{o}rperforschung, Heisenbergstrasse 1,%
\\
D-70569 Stuttgart, Germany\\
2) Industrial Research Ltd., P.O. Box 31310, Lower Hutt,\\
New Zealand }
\date{20.11.99}
\maketitle

\begin{abstract}
From magnetisation measurements we provide evidence that the ferromagnetic
superconductor RuSr$_{2}$GdCu$_{2}$O$_{8}$ with {\it T}$_{c}$=45 K and {\it T%
}$_{M}$=137 K exhibits a sizeable diamagnetic signal at low temperature (%
{\it T}%
\mbox{$<$}%
{\it T}$^{{\it ms}}$=30 K) and low magnetic field ({\it H}$^{ext}$%
\mbox{$<$}%
30 Oe), corresponding to a bulk Meissner-phase. At intermediate
temperatures, {\it T}$^{{\it ms}}$%
\mbox{$<$}%
{\it T}%
\mbox{$<$}%
{\it T}$_{c}$, a spontaneous vortex phase forms which is characterized by
unique thermal hysteresis effects. We argue that a recent negative report
[C.W. Chu et al., cond-mat/9910056] regarding the Meissner-effect in Ru-1212
can be explained by impurity scattering or grain size effects.
\end{abstract}


\bigskip \allowbreak

\allowbreak \noindent Superconductivity (SC) and ferromagnetism (FM) are two
antagonistic phenomena. The question as to whether both order parameters
(OP) can coexist on a microscopic scale has attracted a great deal of
ongoing interest. Experimentally, a coexistence of SC and long-range FM
order was first discovered in 1976 in the ternary rare-earth compounds ErRh$%
_{4}$B$_{4}$ \cite{Fertig1} and HoMo$_{6}$S$_{8}$ \cite{Moncton1}. In these
materials the SC state forms at higher temperature ({\it T}$_{c}$%
\mbox{$<$}%
10 K) than the FM state ({\it T}$_{M}$%
\mbox{$<$}%
1 K), however, both temperatures are rather low. The formation of the FM
state eventually leads to the destruction of SC\ (reentrant behavior).
Albeit, there exists a narrow intermediate temperature range where both SC
and FM order can coexist. In this intermediate state the FM order exhibits a
spiral modulation or a domain-like structure (depending on the magnetic
anisotropy of the system). The modulation of the FM OP helps to circumvent
the detrimental pairing-breaking effect due to the exchange interaction
(EXI), which prevents singlet pairing (but not triplet pairing) by lifting
the degeneracy of the spin-up and spin-down electrons of a Cooper-pair, and
the electromagnetic interaction (EMI), which induces screening currents that
suppress SC once the internal fields exceed the upper critical field {\it H}$%
_{c2}$ $\cite{Bulaevski1}.$ Likewise also the SC OP\ may be spatially
modulated as realized in a spontaneous vortex phase (in response to the EMI)
or in a Fulde-Ferrell-Larkin-Ovchinnikov (FFLO) phase \cite{Fulde1} (in
response to the EXI).

Renewed interest in the interplay between FM and SC order has been
stimulated by the recent discovery of coexistence of FM and SC order in the
ruthenate-cuprate compound RuSr$_{2}$GdCu$_{2}$O$_{8}$ (Ru-1212) \cite
{Bauernf1,Tallon1,Bernhard1}\ (and possibly also in RuSr$_{2}$(Gd,Ce)$_{2}$Cu%
$_{2}$O$_{10}$ (Ru-1222) \cite{Bauernf1,Felner1}). In these materials the FM
transition occurs at a considerably higher temperature than the SC one,
i.e., in Ru-1212 {\it T}$_{M}$=132-138 K and {\it T}$_{c}\approx $45 K.
Rather surprisingly, it was found that the onset of SC does not induce any
significant modification of the FM order \cite{Bernhard1}. On the other
hand, it is still an open question as to how the SC OP, which is thought to
originate in the CuO$_{2}$ bilayers, is modified in the presence of the
already developed FM OP, which involves the moments in the Ru-O layers.
Recent proposals include the possibility of a FFLO-type state \cite{Pickett1}
or of a spontaneous vortex phase (SVP) \cite{Pickett1,Chu1,Felner1}.
Obviously, these new materials with their novel and extraordinary properties
promise to be unique model systems for studying the complex interplay of SC
and FM order.

First of all, however, one has to worry about the chemical and structural
homogeneity of these complex materials. One is confronted with three major
concerns: 1) are the FM and the SC phases intrinsic to the Ru-1212 compound,
or is one of them related to a minor impurity phase; 2) does the FM OP
persist throughout the entire volume of the sample; and 3) is the same true
for the SC OP? Already there exists ample evidence that the answer to the
first two questions is positive. High-resolution synchrotron X-ray
diffraction \cite{Attfield1} and neutron diffraction measurements \cite
{Chmaisem1,Keimer1} indicate a high structural and chemical homogeneity of
our Ru-1212 samples with no detectable impurity phases. Secondly, muon-spin
rotation ($\mu $SR) measurements \cite{Bernhard1}\ and later electron-spin
resonance (ESR) measurements \cite{Fainstein1}\ have shown that the FM order
is a uniform bulk effect. The remaining unresolved third question thus
concerns the homogeneity of the SC phase. Evidence in favor of a bulk SC
state has been obtained for Ru-1212 from specific heat measurements where a
sizeable peak in the specific heat coefficient $\gamma $ was observed at 
{\it T}$_{c},$ comparable to that for non-magnetic underdoped Y-123 or
Bi-2212 cuprates with a similar {\it T}$_{c}\sim $40-50 K \cite{Tallon2}. On
the other hand, Chu and coworkers recently casted doubts as to whether
Ru-1212 is a bulk SC \cite{Chu1}. They find that a bulk Meissner-effect,
generally considered as the key indicator for bulk SC, does not exist in
Ru-1212. They argue that the SC signal might be due to an impurity phase
which is not even detectable by X-ray or neutron diffraction experiments.
Alternatively, they suggest that the absence of a Meissner-effect could be
attributed to the creation of a SVP. Such a SVP can be expected to form in a
FM superconductor if the spontaneous magnetisation, 4$\pi ${\it M}, exceeds
the lower critical field {\it H}$_{c1}$ (as defined in the absence of the
spontaneous magnetisation), i.e., if 4$\pi ${\it M} 
\mbox{$>$}%
{\it H}$_{c1}$({\it T}=0). \cite{Varma1,Pickett1,Chu1} Otherwise, if {\it H}$%
_{c1}$({\it T}=0)%
\mbox{$>$}%
4$\pi ${\it M}, the Meissner-state will be stable at low temperature.
Moreover, since 4$\pi ${\it M} is only weakly {\it T}-dependent below {\it T}%
$_{c}\ll ${\it T}$_{M}$ while {\it H}$_{c1}$(T) falls to zero at {\it T}$%
_{c} $, a transition to an intermediate SVP will occur at the temperature 
{\it T}$^{{\it ms}}$ where {\it H}$_{c1}$({\it T}$^{{\it ms}}$)=4$\pi ${\it M%
}.

In this Letter we present low-field dc magnetisation measurements on
polycrystalline Ru-1212 samples, which provide evidence that a bulk
Meissner-state develops in the pure compound at low temperature, with {\it T}%
$^{ms}\leq 30$ K varying from sample to sample. In addition, we show that
the SVP, which forms at intermediate temperatures {\it T}$^{{\it ms}}<T<$%
{\it T}$_{c},$ is characterized by unique thermal hysteresis effects. We
argue that the absence of a Meissner-phase in Ru-1212 as reported by Chu et
al. \cite{Chu1} can be explained in terms of a moderate reduction of {\it H}$%
_{c1}$ due to impurity scattering or grain size effects.

Two polycrystalline, pure (SC) RuSr$_{2}$GdCu$_{2}$O$_{8}$ samples ({\bf A}
and {\bf B}), and one Zn-substituted (non-SC) RuSr$_{2}$GdCu$_{2.94}$Zn$%
_{0.06}$O$_{8}$ sample ({\bf C}) have been prepared as described previously 
\cite{Tallon1,Bernhard1}.\ The duration and the temperature of the final
sintering step have been slightly varied: 96 h at 1060 $%
{{}^\circ}%
$C in flowing O$_{2}$ for {\bf A} and {\bf C}; and 20 h at 1055 $%
{{}^\circ}%
$C for {\bf B}. It was previously shown that prolonged sintering at 1060 $%
{{}^\circ}%
$C helps to remove 90$%
{{}^\circ}%
$ [100] rotation twins and also a minor degree of intermixing of Ru$%
\leftrightarrow $Cu and Sr$\leftrightarrow $Gd \cite{Tallon1,Bernhard1}.
Apart from these differences, high-resolution X-ray diffraction (XRD) \cite
{Attfield1} and neutron diffraction measurements \cite{Chmaisem1,Keimer1}
have confirmed that our samples contain no impurity phases above the limits
of sensitivity ($\sim $1\%). The electronic properties have been
characterized by resistivity and thermo-electric power (TEP) measurements.
The onset of the drop in resistivity and the temperature where the TEP
becomes zero indicate {\it T}$_{c}=$45 K for samples {\bf A} and {\bf B} and 
{\it T}$_{c}$%
\mbox{$<$}%
4 K for the Zn-substituted sample {\bf C}. \cite{Tallon1,Bernhard1,Tallon2}
All samples have been further investigated by $\mu $SR measurements which
confirm that the FM ordering of the Ru-moments and also the
antiferromagnetic (AF) ordering of the Gd-moments is hardly affected by the
thermal treatment or by Zn-substitution \cite{Bernhard2}. The DC
magnetisation measurements have been performed with a Quantum Design MPMS7
magnetometer.

Figure 1 shows the volume susceptibility, $\chi _{V}$, of samples {\bf A}
and {\bf C} obtained after zero-field cooling (zfc) to 2 K before applying
an external field {\it H}$^{ext}$ = 6.5 Oe. A value of $\sim $95\% of the
ideal density $\rho $\ = 6.7 g/cm$^{3}$ of stoichiometric RuSr$_{2}$GdCu$_{2}
$O$_{8}$ with {\it a} = 3.84 \AA\ and {\it c} = 11.57 \AA\ \cite
{Tallon1,Attfield1} has been determined for samples {\bf A} and {\bf C} and
was used to calculate the susceptibility. We have not corrected for the
demagnetization factor which should be small since the samples have a
bar-shaped form and H$^{ext}$ is parallel to the long axis. The FM ordering
of the Ru-moments is marked for both samples by a cusp in $\chi _{V}$ at 
{\it T}$_{M}$=137 ({\bf A}) and 132 K ({\bf C}). In sample {\bf C,} $\chi
_{V}$ exhibits a pronounced increase below 50 K due to the paramagnetic
contribution of the Gd-moments which order antiferromagnetically (AF) at 2.5
K (as indicated by a cusp in $\chi _{V}$ and as seen in $\mu $SR \cite
{Bernhard1,Bernhard2} and neutron diffraction \cite{Keimer1}). For the SC
sample A, however, a sizeable diamagnetic shift occurs at {\it T}$^{{\it ms}}
$ = 30 K. This is not the thermodynamic SC transition, which occurs at {\it T%
}$_{c}\approx $45 K \cite{Tallon2} and is marked by a weak diamagnetic shift
as shown in the enlargement in the inset to Fig. 1.

Figures 2 shows the field-cooled (fc) volume susceptibility, ${\it \chi }%
_{V} $, of sample {\bf A} (solid lines) at 0.5$\leq H_{ext}\leq $500 Oe and
of sample {\bf C} (dotted lines) at 0.5, 2.5 and 100 Oe. The external field
was changed at 200 K with the sample in the paramagnetic state. The low
fields were measured by a comparison of the respective paramagnetic signals
at 200 K with the signals measured at 50$\leq H_{ext}\leq $500 Oe. For both
samples a spontaneous magnetisation develops at similar temperatures of {\it %
T}$_{M}$=137 K ({\bf A}) and {\it T}$_{M}$=132 K ({\bf C}) and below $\sim $%
100 K it rises almost linearly with decreasing temperature. A clear
difference appears only below 30 K where the susceptibility of the SC sample
is strongly reduced as compared to the Zn-substituted one. For the
Zn-substituted sample $\chi _{V}$ increases rather steeply at low T due to
the paramagnetic contribution of the Gd-moments. In marked contrast, for the
SC sample $\chi _{V}$ decreases suddenly below {\it T}$^{{\it ms}}$=30 K
(corresponding to a sizeable diamagnetic shielding) and remains almost {\it T%
}-independent below {\it T}$^{{\it ms}}$. Evidently, in the SC sample the
paramagnetic Gd-moments are screened against the external field and also the
internal spontaneous magnetisation. In other words, the SC sample is in a
bulk Meissner state at {\it T}%
\mbox{$<$}%
{\it T}$^{{\it ms}}$. Apparently, the paramagnetic Gd-moments provide a very
useful probe for the Meissner-effect. Below we argue that the observed
behavior is indicative of a transition from a Meissner-phase at {\it T}%
\mbox{$<$}%
{\it T}$^{{\it ms}}$=30 K to a SVP at {\it T}$^{{\it ms}}$=30 K%
\mbox{$<$}%
{\it T}%
\mbox{$<$}%
{\it T}$_{c}$=45 K. The volume fraction of the Meissner-phase as estimated
from the size of the diamagnetic shift, $(\chi _{V}(T\rightarrow 0)-\chi
_{V}(T^{ms}))/(\chi _{V}(T^{ms})+1),$ is shown in the inset of Fig. 2(b) as
a function of {\it H}$^{ext}$. Apparently the Meissner-fraction is almost 40
\% at 0.5 Oe but falls very steeply as a function of H$^{ext}$. Our estimate
gives only a lower limit for the Meissner-fraction. The diamagnetic
shielding tends to be reduced by vortex pinning and also by the small
average grain size of around 2-10 $\mu $m (which is further reduced due to 90%
$%
{{}^\circ}%
$ [100] rotation twins and antiphase boundaries) which is almost comparable
to the magnetic penetration depth $\lambda $. Assuming an average grain
radius {\it r}=3 $\mu $m and an effective magnetic penetration depth $%
\lambda _{eff}=\sqrt[3]{\lambda _{ab}^{2}\lambda _{c}}\sim $500 nm, we
obtain from the Shoenberg-formula $\chi /\chi _{o}=1-(3\lambda /r)\coth
(r/\lambda )+3\lambda ^{2}/r^{2}\approx 0.5$ \cite{Christos1}, i.e. a two
times larger Meissner-fraction. Note that $\lambda _{eff}$ $\approx $500 nm
is quite a reasonable assumption since the unique dependence of {\it T}$_{c}$
on ${\it \lambda }$ in underdoped cuprate SC \cite{Uemura1} implies $\lambda
_{ab}\approx $300 nm for {\it T}$_{c}\approx $45 K, whereas $\lambda _{c}$
typically exceeds 2000 nm \cite{Christos1}. Based on these considerations we
conclude that sample {\bf A} exhibits a bulk Meissner-state with the volume
fraction exceeding 40 \%.

This brings us to the interesting question as to why no evidence of a bulk
Meissner-phase has been obtained in a recent study on seemingly similar
Ru-1212 samples \cite{Chu1}. A straightforward explanation is that $H_{c1}$(%
{\it T}=0) is moderately reduced in these samples. As was noted above, if 
{\it H}$_{c1}(T=0)<4\pi {\it M}$, a SVP will be energetically more favorable
than the Meissner-phase even at zero applied field and at zero temperature. 
{\it H}$_{c1}$ may be reduced by pair-breaking due to magnetic (or
non-magnetic) defects causing a reduction of the SC condensate density and a
commensurate enhancement of $\lambda $ (which is particularly strong in case
of a SC OP with d-wave symmetry \cite{Bernhard3}). We speculate that such
defects may arise, for example, due to some intermixing between Cu and Ru or
to antiphase boundaries in the rotation pattern of the RuO$_{6}$ octahedra
such as observed by x-ray and neutron diffraction \cite
{Tallon1,Attfield1,Chmaisem1}. Also, since the effective value of {\it H}$%
_{c1}$ depends on the ratio of $\lambda $/{\it r}, the morphology of a given
Ru-1212 sample (for example the amount of [100] rotation-twins \cite{Tallon1}%
) may actually determine whether or not it exhibits a Meissner-effect. The
data in Fig. 2 imply that {\it H}$_{c1}$(T=0$)$ in our sample {\bf A}
exceeds 4$\pi ${\it M} by less than 30 Oe since the diamagnetic shift at 
{\it T}$^{{\it ms}}$ diminishes very rapidly as {\it H}$^{ext}$ increases.
At 35 Oe the susceptibility already starts to exhibit a slight paramagnetic 
{\it T}-dependence due to the Gd-moments that are no longer screened against
the local fields. From the remanent magnetisation found after high field
saturation measurements \cite{Bernhard1} we estimate that 4$\pi M$ is of the
order of 50-70 Oe. Note that 4$\pi $M is about 10 times smaller than the
internal field of $\sim $700 Oe as obtained from $\mu $SR measurements \cite
{Bernhard1} or deduced for the case that the Ru-moments of size 1$\mu _{B}$
exhibit purely ferromagnetic order. This difference may indicate that the
Ru-moments exhibit a canted antiferromagnetic order with only a $\sim $10 \%
ferromagnetic component. Under the assumption that 4$\pi ${\it M} is only
weakly {\it T}-dependent below {\it T}$_{c}$%
\mbox{$<$}%
\mbox{$<$}%
{\it T}$_{M}$ and using {\it H}$_{c1}$({\it T}$^{{\it ms}}$)={\it H}$_{c1}$%
(T=0)$\times (1-(T^{ms}/T_{c})^{2})$=4$\pi ${\it M} with {\it T}$^{{\it ms}}$%
/{\it T}$_{c}$=30/45, we then obtain {\it H}$_{c1}$(T=0) of the order of
80-120 Oe. In turn this gives $\lambda =\sqrt{\Phi _{o}/H_{c1}}\sim $400-500
nm in reasonable agreement with our above estimates.

The finding that T$^{ms}$ appears to decrease only slightly from 30 K at 0.5
Oe to 27 K at 10 Oe can be understood due to the random orientation of the
spontaneous magnetisation of the individual FM domains with respect to the
external magnetic field. For very small external fields the domains will not
be aligned and in most domains the effective internal field will be only
marginally increased or even be decreased. However, once these domains
become aligned by a sufficiently large field {\it H}$^{ext}$, the internal
field will suddenly be increased to a value 4$\pi M\ $+{\it H}$^{ext}$%
\mbox{$>$}%
{\it H}$_{c1}$({\it T}=0). For the individual domains the alignment thus
will trigger a sudden transition from a state with a Meissner-phase below 
{\it T}$^{{\it ms}}\simeq $ 30 K to one where the SVP persists to the lowest
temperatures.

In the following we show that the transition temperature of the
Meissner-phase {\it T}$^{{\it ms}}$ varies considerably even among samples
that have been prepared under similar conditions. Figure 3 shows the fc data
at 6.5 Oe for sample {\bf B} which has been sintered at slightly lower
temperature and for a shorter period as described above. Sample {\bf B} has
the same critical temperature {\it T}$_{c}=45K$ as sample {\bf A} (as
confirmed by transport and thermodynamic measurements \cite{Tallon1,Tallon2}%
), but a Meissner-phase forms only at significantly lower temperature {\it T}%
$^{{\it ms}}\approx $16 K. Another interesting feature is the strong thermal
hysteresis of $\chi _{V}$ at the transition from the vortex phase to the
Meissner phase. Upon cooling (solid line) the transition occurs at a
distinctively lower temperature (of about 1 K) than upon warming (dotted
line). Notably, the hysteresis occurs only after the sample has been cooled
below {\it T}$^{{\it ms}}$. It is absent if the sample is only cooled to 
{\it T}=17 K (crosses) and subsequently warmed (open circles). This kind of
hysteresis, in particular the undercooling effect, is indicative of a
first-order transition such as from a SVP to a Meissner-state where the
magnetisation exhibits a discontinuous change. In the SVP flux-lines are
formed which penetrate the sample volume completely. Below {\it T}$^{{\it ms}%
}$, as the Meissner-state develops, the flux-lines are expelled from the
interior of the grains. However, pinning of the vortices by any kind of
defects will lead to an incomplete expulsion of the vortices and thus will
reduce the diamagnetic shift. On warming the sample again above {\it T}$^{%
{\it ms}}$, the flux-lines have to reenter the individual grains. Pinning
will hinder the vortices from reentering the superconducting grains. This
leads to hysteresis as shown in Fig. 3, where the magnetisation upon cooling
is higher than that upon subsequent warming.

In sample {\bf A} the signature of the hysteretic transition at {\it T}$%
^{ms} $ is less pronounced, probably because it contains fewer defects that
act as pinning centers and its transition temperature, {\it T}$^{{\it ms}}$%
=30 K, is almost twice as high. However, yet another kind of thermal
hysteresis related to the magnetisation of the Gd-moments occurs for {\it H}$%
^{ext}\geq $35 Oe, once the SVP persists to low T. As noted above, the
paramagnetic Gd-moments eventually become partially aligned in the local
field at low T and therefore give rise to a sizeable enhancement of the
spontaneous magnetisation. Note, that in the SVP the density of the vortices
is not only determined by {\it H}$^{ext}$ but, in addition, by the
spontaneous magnetisation. Therefore, even though {\it H}$^{ext}$ is
constant for a fc curve, the vortex density tends to increase upon
decreasing the temperature as the Gd-moments become aligned by the local
field. Any vortex pinning thus will lead to thermal hysteresis such as shown
in Fig. 4 where $\chi _{V}$ is lower upon cooling (solid lines) than upon
warming (dotted lines). We have confirmed that such hysteretic behavior of
the magnetisation does not occur for the Zn-substituted (non-SC) sample C
(not shown here). The observed unique hysteretic behavior therefore gives
yet another demonstration of the direct interaction between SC and FM order
in the SVP and thus of their microscopic coexistence.

In summary, we have presented dc magnetisation measurements which provide
evidence that the FM superconductor RuSr$_{2}$GdCu$_{2}$O$_{8}$ develops a
bulk Meissner-state. In addition, we show that the spontaneous vortex phase
forming at intermediate temperature, {\it T}$^{{\it ms}}<T<${\it T}$_{c},$
is characterized by unique thermal hysteresis effects. We outline that the
absence of a Meissner-phase in Ru-1212 as reported by Chu et al. \cite{Chu1}
can be explained in terms of a moderate reduction of {\it H}$_{c1}$ due to
impurity scattering or grain size effects.

C.B. acknowledges discussion with T. Holden, C. Niedermayer and C.
Panagopoulos.

\section{\protect\bigskip {\bf Figure Captions}}

Figure 1: Zero-field-cooled (zfc) volume susceptibility, ${\it \chi }_{V,}$
at 6.5 Oe of the pure sample {\bf A} (solid line) and the Zn-substituted
sample {\bf C} (dotted line). Inset: Susceptibility of sample {\bf A} around
the SC transition, {\it T}$_{c}=45$ K, shown on an enlarged scale.

\bigskip

Figure 2: Field-cooled (fc) volume susceptibility, ${\it \chi }_{V},$ (a) of
the pure sample {\bf A} at 2.5, 6.5, 10, 20, 35, 50, 100 and 500 Oe (solid
lines) and the Zn-substituted sample {\bf C} at 2.5 and 100 Oe (dotted
lines); (b) of sample {\bf A} at 2.5, 1.5, 0.75 and 0.5 Oe (solid lines) and
sample {\bf C} at 0.5 Oe (dotted line).

\bigskip

Figure 3: Low temperature fc curve at 6.5 Oe for the pure sample {\bf B}
which has the same {\it T}$_{c}=45K$ as sample {\bf A} but has been prepared
under slightly different conditions as noted in the text. The Meissner-phase
forms at significantly lower temperature {\it T}$^{{\it ms}}\approx $16 K.
Note the thermal hysteresis of ${\it \chi }_{V}$ around {\it T}$^{{\it ms}}$
which is absent if the sample is only cooled to {\it T}=17 K (crosses) and
subsequently warmed (open circles).

\bigskip

Figure 4: Thermal hysteresis of the fc data of sample {\bf A} at 35, 50,
100, 250 and 500 Oe. The solid lines (dotted lines) show $\chi _{V}$ upon
cooling (warming). Arrows indicate the direction of the temperature change.
At 100 Oe two hysteresis curves for cooling to 2 and 4 K are shown by the
thick and thin dotted lines, respectively.

\end{document}